# Efficient interlayer interactions in non-planar molecule/transition-metal dichalcogenide two-dimensional heterostructure

*Johann Alvarez[1†], Kamal Harrison[1†], John Koptur-Palenchar[1], Miguel Gakiya-Teruya[2], Kenji Watanabe[3], Takashi Taniguchi[4], Michael Shatruk[2,5], Xiao-Xiao Zhang[1]**

AUTHOR ADDRESS

[1] Department of Physics, University of Florida, Gainesville, Florida, 32611, USA

[2] Department of Chemistry and Biochemistry, Florida State University, Tallahassee, Florida 32306, USA

[3] Research Center for Electronic and Optical Materials, National Institute for Materials Science, 1-1 Namiki, Tsukuba 305-0044, Japan

[4] Research Center for Materials Nanoarchitectonics, National Institute for Materials Science, 1-1 Namiki, Tsukuba 305-0044, Japan
[5] FSU Quantum Initiative, Florida State University, Tallahassee, Florida 32310, USA

*[†]J.A. and K.H. contributes equally to this work*

**Corresponding author: xxzhang@ufl.edu*

ABSTRACT

Integration of molecules and two-dimensional (2D) atomic materials offer new opportunities for hybrid electronic devices. Most progress has focused on interfacing planar molecule. In contrast, incorporating non-planar molecules into 2D heterostructures remains challenging due to the three-dimensional orientations of ligands. Here we investigate interlayer interactions between discrete non-planar spin-crossover (SCO) molecules, $[Fe(tBu_2qsal)_2]$ (FTBQS), monolayer transition metal dichalcogenides (TMDs), and hexagonal boron nitride (hBN) spacers through optical spectroscopies. FTBQS features bulk molecular crystal layers and can be mechanically exfoliated, enabling deterministic assembly of clean hybrid heterostructures. Robust photoluminescence (PL) quenching, even with an hBN spacer, indicates efficient Förster resonance energy transfer with a long decay distance. Gate-dependent optical spectroscopy further identifies charge transfer from the doping shifts in the TMD layers due to the TMD/FTBQS electronic band alignment. Our results establish efficient interlayer interactions between non-plana, discrete molecules and 2D crystals for hybrid electrical, optical, and spintronic devices.

Main Text:

Two-dimensional (2D) hybrid heterostructures that integrate molecules and 2D atomic materials have attracted growing interest[1,2] and represent an emerging platform for exploring next-generation functional devices. Molecule-based materials offer chemically tunable electronic, optical, and spin properties and can exhibit phase transitions under external stimuli, including thermal, optical, and electrical excitations[3-7]. By combining these molecular functionalities, 2D hybrid heterostructures can serve as building blocks for nanoscale molecular devices or provide molecular switching capabilities for 2D atomic materials, offering applications in sensing, energy harvesting, optoelectronics, and quantum computing[8,9].

Previously studied hybrid 2D interfaces primarily involved planar molecules, such as pentacene, tetrathiafulvalene, and phthalocyanine, for their transistor responses and optical couplings [10-15]. Their molecular geometries favor a close proximity to 2D crystals and facilitate strong and homogeneous interlayer interactions. On the other hand, non-planar molecules often offer precise control of electronic and optical properties and host complex functionalities including chirality[16-18], molecular magnetism[19-21 22-24] , and formation of supramolecular architectures[25,26]. However, it remains unclear if similarly efficient interlayer interactions can be achieved with discrete, non-planar molecules, where extended ligands and three-dimensional molecular geometries can substantially increase the interlayer distance. In addition, it is challenging to integrate them with 2D atomic materials or planar surfaces, as conventional deposition methods tend to introduce thickness and molecular orientation inhomogeneity[27,28].

To exploit the potential of hybrid 2D heterostructures with non-planar molecules, it is important to understand the different types of interlayer interactions across length scales in such systems. Charge transfer and Dexter-type energy transfer require substantial wave-function overlap and are significant only at short-range distances. On the other hand, Coulomb-mediated processes, mainly Förster resonance energy transfer (FRET) and dielectric screening effects, can extend beyond a few nanometers and provide long-range interlayer interactions. These interlayer interactions essentially determine the electronic and optical properties of hybrid heterostructures and can give rise to emergent interfacial phenomena. Previous demonstrations[9,12 10,13-15,29 30 31-33] include hybrid *p-n* junctions, optoelectronic prototypical devices, and spin-dependent sensing with magnetic molecules. Establishing efficient and controllable interlayer coupling between non-planar molecules and 2D materials can open new avenues for exploring emergent physical properties and expanding device functionalities in low-dimensional systems.

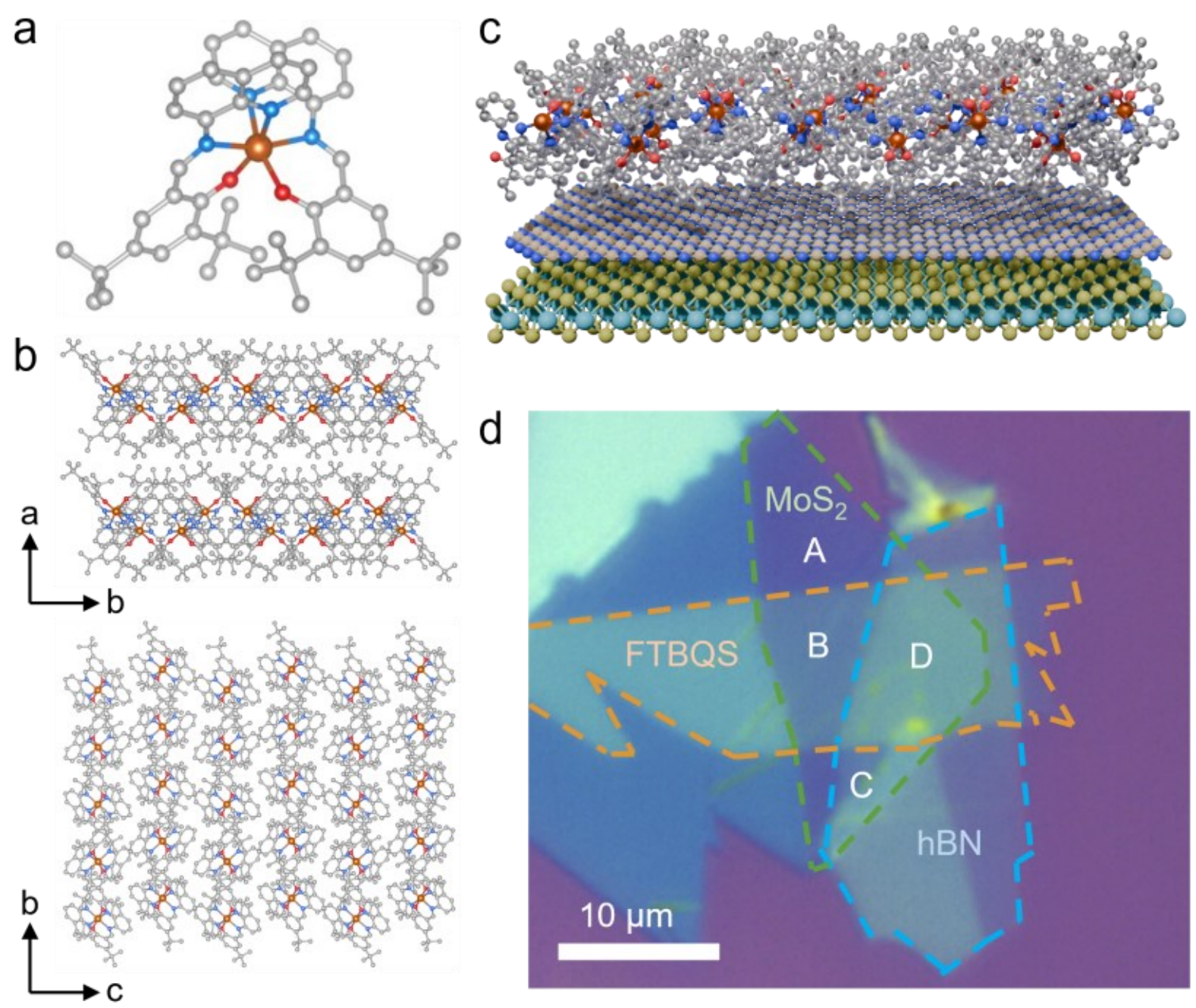

Figure 1. (a) Atomic structure of the [Fe(tBu$_2$qsal)$_2$] (FTBQS) molecule with the H atoms omitted for clarity. Orange: Fe; Blue: N; Grey: Carbon; Red: Oxygen. (b) Crystal packing of FTBQS along the a and c axes. (c) Schematic of a representative 2D heterostructure with (from top to bottom) FTBQS, hBN, and $MoS_2$.Reduce hBN to monolayer (d) Optical micrograph of the dry-transfer stacked 2D heterostructure. A-D labels the 1L $MoS_2$, FTBQS/$MoS_2$, hBN/$MoS_2$, and FTBQS/hBN/$MoS_2$ regions, respectively.

Here we investigate the interlayer interactions between non-planar spin-crossover (SCO) molecules $[Fe(tBu_2qsal)_2]$, hereafter abbreviated as FTBQS[34,35], and monolayer transition-metal dichalcogenides (TMDs). Although the individual FTBQS molecules are non-planar (Fig. 1a), their layered molecular crystal packing (Fig. 1b) permits mechanical exfoliation of single-crystalline flakes with lateral sizes of tenths of micrometers and thicknesses of a few molecular layers [31]. Compared with films conventionally obtained from thermal deposition or drop casting, these mechanically exfoliated molecular flakes exhibit well-defined thicknesses and substantially reduced structural inhomogeneity. Their single-crystalline structure also enables cooperative SCO transitions to persist down to the few-layer limit, as demonstrated previously[31], opening pathways to study stimulus-responsive SCO behaviors in molecule/solid-state heterostructures. Using deterministic dry-transfer methods[34], we can further manipulate these molecular layers and fabricate homogeneous and clean hybrid heterostructures with minimal thermal or chemical degradation.

To investigate the interlayer interactions at different length scales, we use additional thin hBN layers as spacers between the FTBQS and TMD layers. By directly comparing the optical responses of adjacent regions with and without the spacer within the same heterostructure, we can separate and characterize the short-range and long-range interactions explicitly. We observed efficient interlayer coupling, manifested by modified photoluminescence (PL) and reflectivity spectra of the TMDs in the heterostructure regions. In particular, strong PL quenching persists with hBN spacers thicker than 10 nm, indicating long-range nonradiative energy transfer that we attribute to FRET. PL quenching efficiencies of different spacer thicknesses can be fitted to a nonradiative decay that follows a $\sim z^{-4}$ as a function of interlayer distance $z$, consistent with the expected scaling of FRET in 2D systems[36-38]. Furthermore, the quenching efficiencies across different TMDs show qualitative agreement with the expected trends based on the spectral overlap between TMD donor emission and FTBQS acceptor absorption. We also identify short-range charge transfer from doping-level shifts in the TMD layer, as measured by gate-dependent PL. Opposite doping trends were observed for different TMD compounds, consistent with their different electronic level alignments with FTBQS and suggesting a type-II band alignment. Our work establishes the relevant interaction length scale in this prototypical non-planar molecule/2D heterostructure. The results demonstrate efficient long-range FRET and short-range charge transfer despite the extended molecular geometry, which can serve as promising coupling mechanisms for non-planar molecule sensing and switching for future device applications.

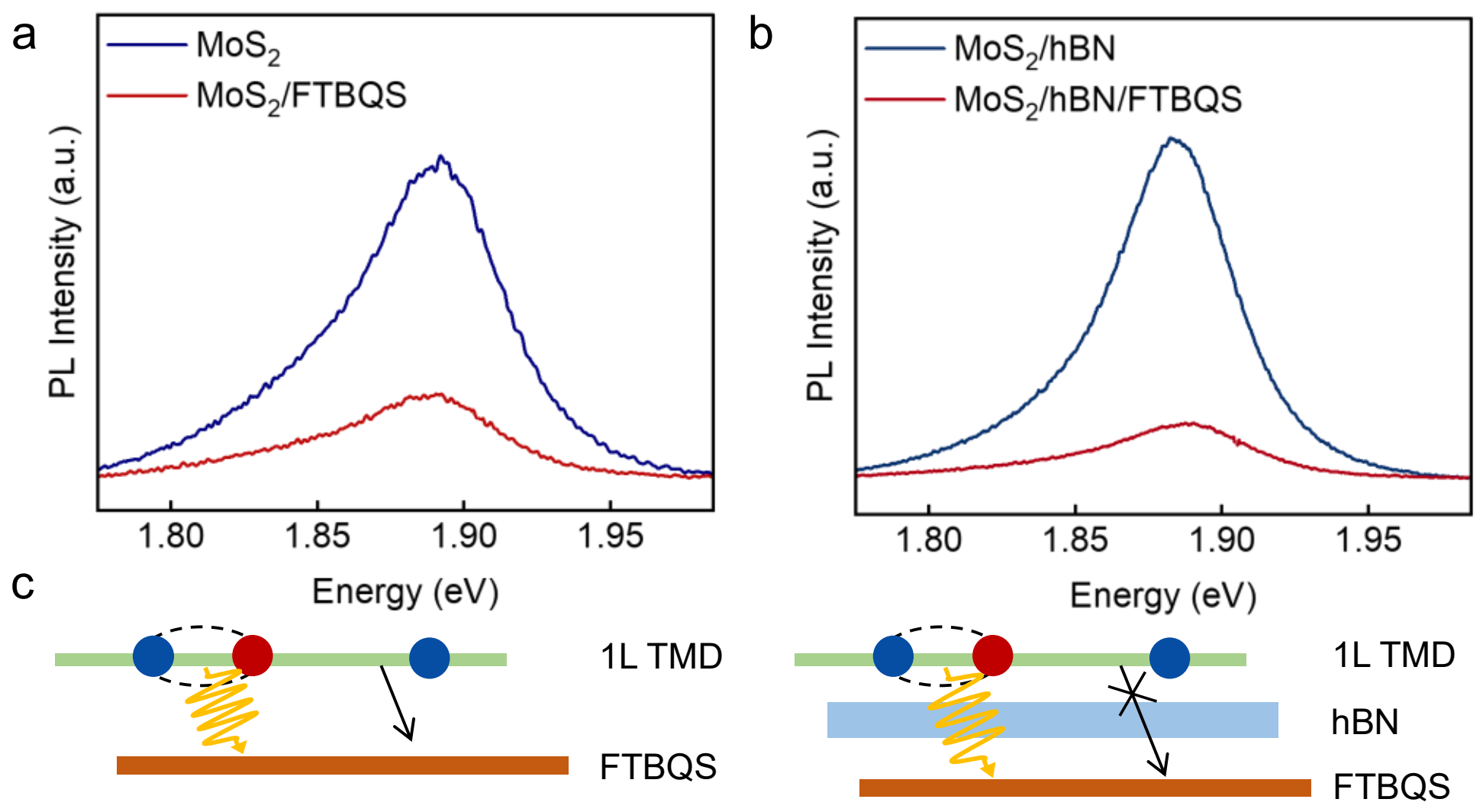

Figure 2. PL emission spectra for (a) $MoS_2$ and $MoS_2$/FTBQS, and (b) $MoS_2$/hBN and $MoS_2$/hBN/FTBQS heterostructures. (c) Heterostructure schematic of TMD/FTBQS and TMD/hBN/FTBQS. The latter illustrates dipole-dipole interactions between the 1L TMD and FTBQS flakes while preventing charge transfer across the thin dielectric gap (hBN) while allowing long-range dipole-dipole interactions to occur. Measurements are taken at room temperature.

The FTBQS molecule (Fig. 1a) contains tert-butyl ($^t$Bu) groups that interact weakly with their neighbors and quinoline groups that interact strongly via non-covalent π- π and π-σ contacts[34], which results in layered crystal packing (Fig. 1b) with interlayer van-der-Waals interactions between the $^t$Bu groups. Molecular layers form perpendicular to the [100] direction, with a single-layer thickness of 1.7 nm. These layers can be mechanically exfoliated onto planar substrates[35]. FTBQS molecules exhibit a spin-bistable transition between the high-spin (HS, paramagnetic) and the low-spin (LS, diamagnetic) states under external stimuli, including temperature, strain, and light exposure. In bulk crystals, the thermal SCO occurs at 117 K on cooling and 129 K on heating, accompanied by changes in the dielectric function and molecular size[34,35].

To assemble the heterostructures, bulk crystals of FTBQS, $MoS_2$, $MoSe_2$, $WS_2$, $WSe_2$, and hBN were mechanically exfoliated onto $SiO_2$/Si and PDMS substrates using the scotch-tape method, and the exfoliated thin layers were identified by their contrast under a white-light optical microscope. The TMD/FTBQS heterostructures were assembled using the PDMS dry-transfer method for a systematic study of their interlayer interactions, as schematically shown in Fig. 1c. The deterministic nature of this heterostructure transfer allows us to compare and isolate responses from the heterostructures. Figure 1d shows a white-light optical microscope image of a $MoS_2$/hBN/FTBQS heterostructure. The heterostructure includes a few-layer FTBQS flake stacked atop a monolayer TMD, with a few-layer hBN flake between them, providing adjacent regions of 1L $MoS_2$, FTBQS/$MoS_2$, hBN/$MoS_2$, and FTBQS/hBN/$MoS_2$ for independent study. The heterostructure area with the additional thin hBN layer serves as an insulating spacer.

We identify strong interlayer interactions optically through PL spectroscopy. Figure 2a compares the room-temperature PL emission from the region of monolayer (1L) $MoS_2$ (region A of Fig. 1d) and 1L $MoS_2$/FTBQS (region D of Fig. 1d) heterostructure on the same exfoliated $MoS_2$ flake. The PL peak in monolayer $MoS_2$ is around 1.89 eV, consistent with previous reports of the A exciton emission[39,40]. The emission peak energy remains the same for $MoS_2$ with the addition of FTBQS, indicating no substantial strain change[41] with the heterostructure fabrication. Most notably, we observed significant PL quenching in the heterostructure region, indicating efficient interlayer coupling. To identify the origin of the PL quenching, we investigated the same heterostructure with a thin hBN (~ 3 nm) spacer. The added hBN layer prevents short-range interactions such as charge transfer and strain effects, while allowing long-range interactions like dipole-dipole interactions at the molecular/solid-state interface (Fig. 2c). This spacer layer also serves to eliminate extrinsic factors for PL quenching, including potential surface contamination and degradation at the heterostructure interface.

Figure 2b compares the PL spectra of the $MoS_2$/hBN (region C of Fig. 1d) and $MoS_2$/hBN/FTBQS (region D of Fig. 1d) regions at room temperature, which still have substantial quenching over the heterostructure. Supplementary Fig. S4 shows the consistent quenching at low temperature. The results from the sample with the hBN spacer show that the quenching comes from long-range nonradiative channels instead of charge transfer or short-range energy transfer. We attribute this to arise from FRET process. In this heterostructure, the donor is the excitons in the TMD layer, and the acceptor is the dipole moments in FTBQS molecules[34]. During FRET, excitons (and therefore donor dipoles) in TMDs undergo nonradiative decay, transferring their dipole moment to the FTBQS layer, leading to PL quenching. When comparing Fig. 2a and 2b, the quenching is even more efficient with the additional hBN spacer, despite the slower FRET rate with the increased donor-acceptor distance. This result can also be understood by considering the differences in $MoS_2$ exciton lifetimes when placed on the Si/$SiO_2$ substrate and hBN. Exciton lifetime increases for $MoS_2$ on hBN as a result of a decrease in the nonradiative relaxation rate due to reduced substrate interactions. The competing FRET nonradiative process can therefore lead to more efficient exciton quenching.

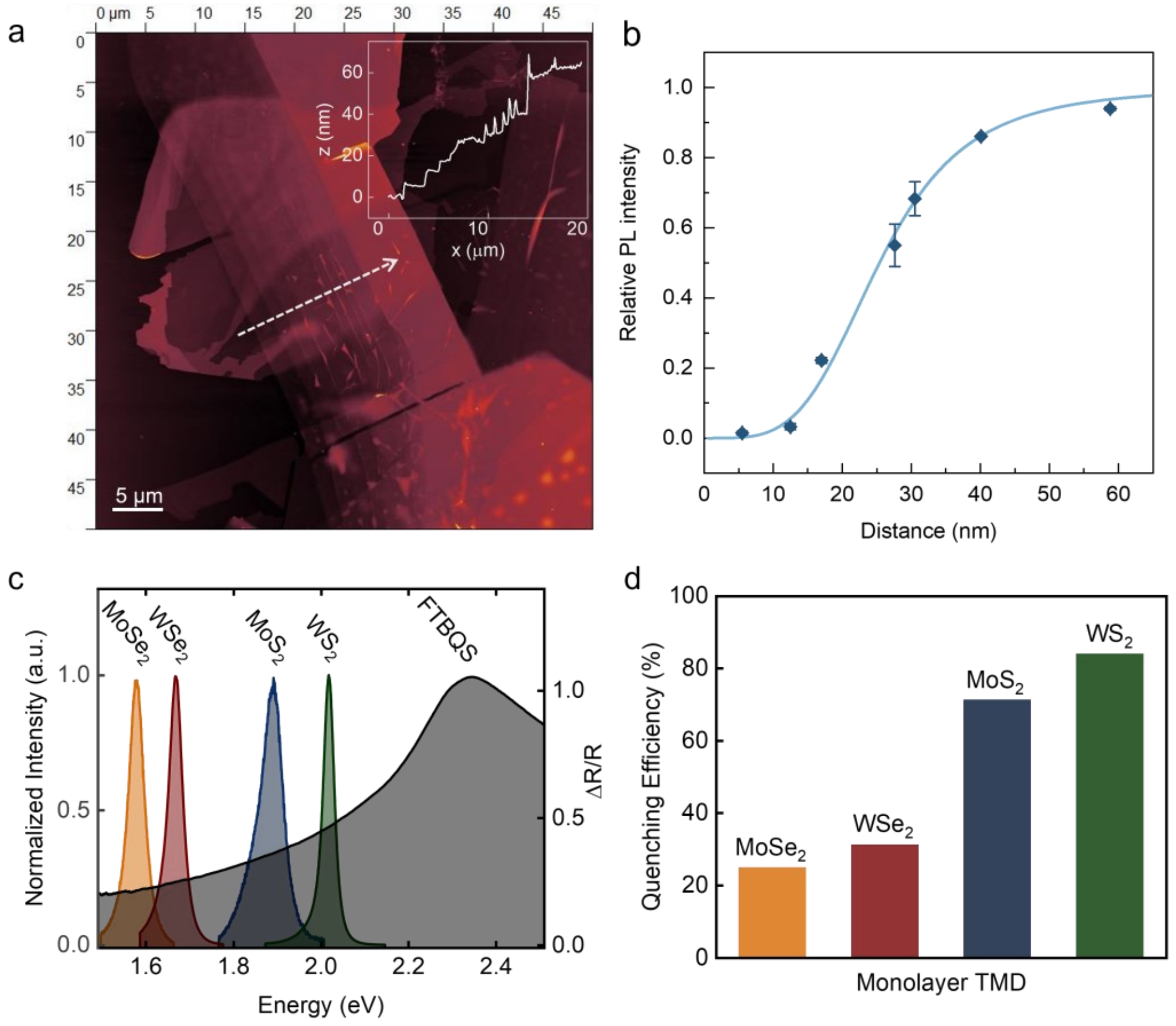

Figure 3. (a) AFM of a 1L $MoS_2$/hBN/FTBQS device, where hBN has multiplayer steps ranging from 5-60nm. The inset plots the height linecut along the white dashed line. (b) The relative PL intensity ploted as a function of the hBN heights. The solid line corresponds to fitting with FRET rate scaling of $z^{-4}$. The error bars correspond uncertainties in the fitted quenching amplitude ratio. (c) Normalized PL intensities of the four TMDs and overlayed with the reflection contrast, which is proportional to the absorption spectrum of few-layer FTBQS at room temperature. The FTBQS reflectivity is taken with thin FTBQS exfoliated on transparent fused-silica. (d) Measured PL quenching efficiencies of (left to right) $MoSe_2$, $WSe_2$, $MoS_2$, and $WS_2$ heterostructures with FTBQS. All data here are taken at room temperature and under linear excitation conditions. Individual spectra are included in Fig. S3. Additional device data can be found in the Supplementary.

To investigate the distance dependence and verify the nature of FRET, we prepared a sample with a hBN spacer of varying thicknesses between FTBQS and monolayer $MoS_2$. Figure 3a shows the atomic force microscopy (AFM) image of the heterostructure, and supplementary Fig. S2 shows the white-light microscope image with corresponding layer outlines. The hBN spacer provides varying distances from 5 nm to 60 nm, as shown in the insert of Fig. 3a, allowing us to compare PL quenching under otherwise identical sample conditions. We plot the relative PL intensity, $A_{Het}/A_{MoS_2}$ comparing the integrated PL intensity from $MoS_2$/hBN/FTBQS and $MoS_2$/hBN in Fig. 3b. The quenching decreases with increasing spacer distance, and with thick hBN above 30 nm, the quenching becomes insignificant. We fit the distance dependence of relative PL intensity to $\left[1 + \frac{\gamma_F(z)}{\gamma_R}\right]^{-1}$, where $\gamma_R$ is the radiative recombination rate and $\gamma_R$ is the FRET decay rate as a function of interlayer distance $z$, assuming other nonradiative rates remain the same. As shown in Fig. 3b, the data fits well with $\gamma_F \propto z^{-4}$, which is the expected scaling relationship for FRET[36-38] between 2D arrays of donor dipoles (monolayer TMD) and acceptor dipoles (finite-thickness molecular crystal) . Notably, the FRET has a long decay distance and remains efficient even above 20 nm, based on our measured distance dependence. This is due to the slower decay scaling for dipoles in 2D layers. In addition, we note that the TMD exciton dipoles have relatively large spatial extent, with the ground state of ~ 2 nm and the first excited state of ~ 5-8 nm[42]. The larger spatial extent of the donor TMD dipoles can also contribute to the long FRET decay range.

The FRET rate is determined by the overlap between the donor's emission and the acceptor's absorption spectrum[36,43]. Here we can verify such FRET rate variation by examining the PL quenching efficiencies of different TMD heterostructures. As shown in Fig. 3c, among the PL spectra of different TMD compounds, $WS_2$ and $MoS_2$ exhibit a higher overlap with the FTBQS absorption spectrum (dark gray area). The FTBQS spectrum is taken with reflection contrast on a transparent substrate. With Kramers-Kronig relations at thin-film limit, it is proportional to the imaginary dielectric function that can be used in FRET spectral overlap estimate[44]. To assess the quenching effect of TMD/FTBQS structures, we extract and analyze their quenching efficiency (QE), calculated as QE = $1 - \frac{A_{Het}}{A_{TMD}}$, where $A_{TMD}$ and $A_{Het}$ are the integrated intensity of the PL spectra in the TMD and TMD/FTBQS heterostructures, respectively. Figure 3d shows the calculated QE of TMD/FTBQS heterostructures at room temperature. The corresponding PL quenching spectra are shown in Fig.S3. The changes in QE in different TMD monolayers show good correlation with the overlapping emission/absorption in Fig. 3d, where the selenide compounds have lower-energy photon emission that appears at the absorption tail of FTBQS molecules. A more quantitative comparison of the spectral overlap and FRET rate will require further information on the radiative rates of the measured TMD samples and is beyond the scope of this work.

Given the short room-temperature exciton lifetime (<10 ps) of monolayer $MoS_2$ on $SiO_2$/Si substrate[45] and the observed efficient quenching, we expect the FRET rate to be on the same order of magnitude as the exciton lifetime. By comparing the $MoS_2$ A and B exciton intensity ratio and their corresponding PL quenching ratios in another device (Fig. S1), we can provide an estimated timescale of this energy transfer. Unlike A excitons, which can have varying exciton lifetimes under different sample conditions, the higher-energy B exciton lifetime corresponds approximately to the ultrafast hot carrier relaxation time and is less than 2 ps[46], while having similar oscillator strength and radiative lifetime as A excitons. Using the A/B exciton intensity ratio, we can estimate the A exciton lifetime in the $MoS_2$/hNB device and the reduced A exciton lifetime in the $MoS_2$/hBN/FTBQS area. Considering the additional nonradiative rates to be the FRET rate, we can obtain an upper bound estimate of the FRET timescale to be 37 ps.

The PL linewidth broadening in the heterostructure region provides estimates of additional interlayer relaxation channels and the energy-transfer rate. We compare PL linewidths of $WSe_2$, which show more symmetric line shapes and overall lower defect-related scattering than $MoS_2$. As can be seen in Fig. 4c, the heterostructure region shows a full-width half-maximum (FWHM) increase of ~10 meV, indicating the opening of a substantial nonradiative scattering and decay channel. The lifetime of the scattering process can be estimated[47] from the spectral width $\Delta\Gamma$ as $\tau = \hbar/\Delta\Gamma$, which yields a decay lifetime ~ 70 fs. We note that this FWHM increase includes both Lorentzian and Gaussian width increases, with contributsion from both homogeneous and inhomogeneous broadening. This new decay channel arises from the additional FRET process, as well as other possible extrinsic factors, including unintentional strain and surface dopants during heterostructure fabrication. We can therefore infer a lower bound estimate of FRET time to be ~70 fs.

We investigate the temperature dependence of the PL quenching efficiency, in particular to identify whether the SCO transition affects the FRET process. To this end, we measured PL spectra of $WSe_2$/hBN and $WSe_2$/hBN/FTBQS heterostructures from 4 K to 300 K. The presence of the hBN spacer ensures that long-range FRET is the main contributing factor in interlayer interactions. The sharp peak width in $WSe_2$ enables easy comparison of FRET effects on different exciton species. Figure 4a shows the comparison of the PL spectra at 4 K and 300 K. At 4 K, the neutral exciton peak (upper panel of Fig. 4a) is the highest-energy peak near 1.73 eV. Additional emission features are observed at lower energies (~1.65 eV), which are often associated with phonon replicas and defect-related trap states in $WSe_2$[48]. We note that, while the neutral exciton shows PL quenching and linewidth broadening due to FRET, the defect-related states exhibit similar amplitudes across the FTBQS region. This can be understood by considering the differences in dipole amplitude and localization of these defect-trapped states and their phonon replicas. In particular, many of these defect states are charge-related, and the hBN spacer layer prevents charge transfer between the two flakes. The exciton peak at the molecular heterostructure shows a consistent slight blueshift of < 5 meV compared to the monolayer $WSe_2$ region, which can be attributed to the different dielectric screening.

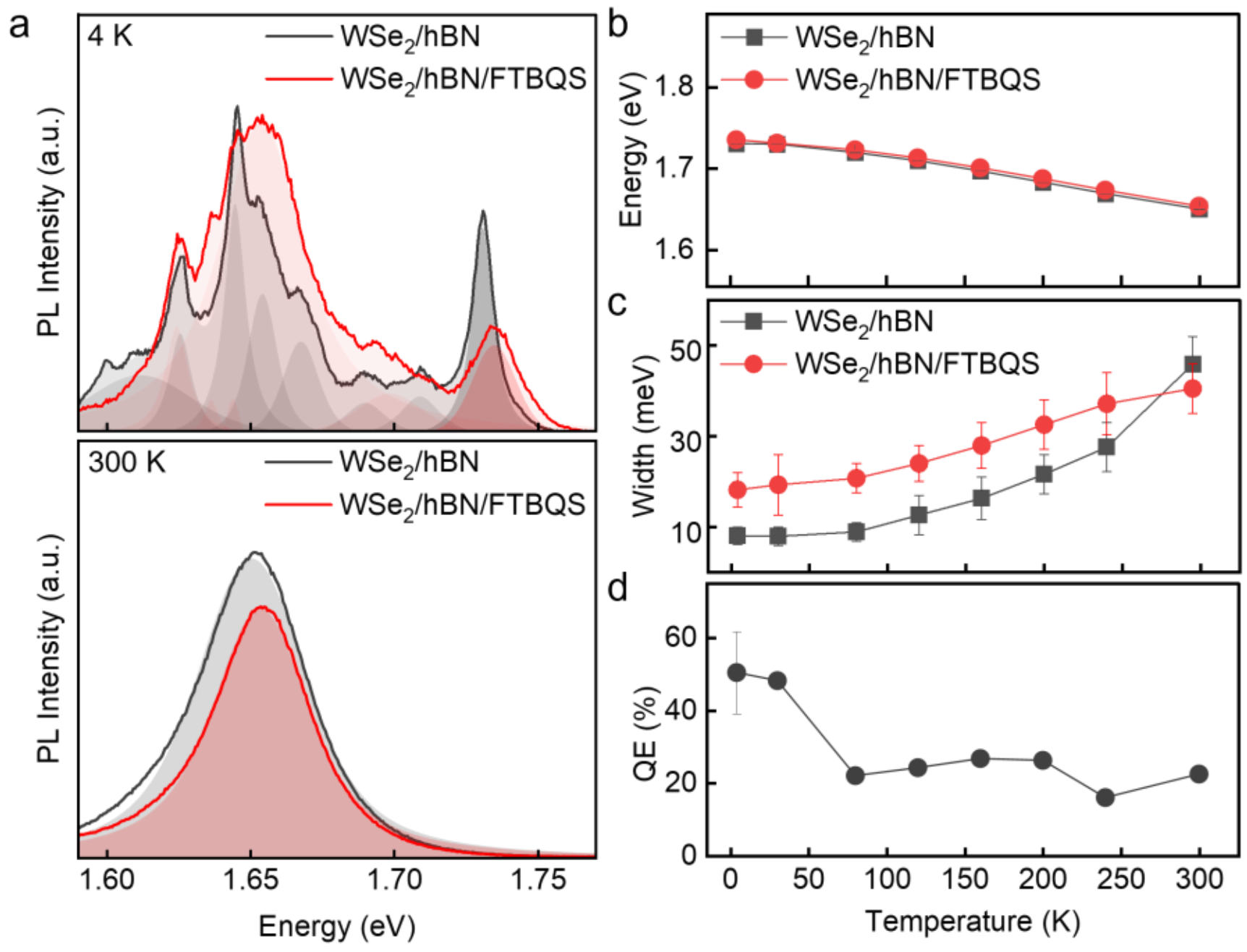

Figure 4. (a) PL spectra (solid line) and Voigt fittings (shaded area) of $WSe_2$/hBN (gray) and $WSe_2$/hBN/FTBQS (red) heterostructure region at 4 K and 300 K. Temperature dependence of (b) neutral exciton energy and (c) peak width for $WSe_2$/hBN (gray) and $WSe_2$/hBN/FTBQS (red) heterostructures. (d) Temperature dependence of the extracted quenching efficiency of neutral exciton.

The temperature dependence of the neutral exciton is summarized in Fig. 4 b- d. The SCO transition is expected near 117 K (bulk), leading to ~2% changes in the in-plane b-axis and in the LUMO-HOMO and optical bandgap[34,35]. In our prior experiments, we have verified that the exfoliated FTBQS flakes also show SCO with a transition temperature (upon cooling) similar to that of the bulk materials[35]. Figure 4d shows that the QE does not exhibit abrupt changes near the expected temperature range of the SCO transition, indicating that the FRET rate in this heterostructure does not show significant changes with the transition. We note that the optical bandgap in FTBQS changes to ~2.3 eV due to the SCO to the low-temperature LS state[49]. This bandgap is significantly higher than and off-resonance compared to the tested TMD donor energies (Fig. 3b). The slight increase in QE below 80 K is likely to arise from the spectral separation of neutral exciton and other defect- and charge-related states, which show minimal quenching (see Fig. 4a, upper panel), and is not intrinsically related to the SCO transition. The linewidth broadening in the heterostructure also does not show significant changes across the expected SCO transition (Fig. 4c), indicating no significant changes in nonradiative relaxation during the phase transition. However, with the demonstrated efficient interlayer energy transfer coupling, we expect that such energy transfer can be used to sense the SCO transition with donor materials that have energy resonant with the HS or LS state optical bandgap.

We also examine the effects of changes in molecular size across the SCO transition in these TMD/FTBQS heterostructures. The heterostructure comparison without the hBN spacer, i.s., the $WSe_2$/FTBQS and $WSe_2$, is expected to serve as a direct probe of strain changes in these two layers. The temperature dependence is shown in Supplementary Fig. S5. Prior strain-dependent studies in TMDs have established correlations between exciton energies and material strain percentage[41]. The expected in-plane size changes in FTBQS can induce strain in the adjacent $WSe_2$ layer and give rise to an abrupt exciton energy shift near the SCO phase transition temperatures[32]. However, as shown in Fig. S2d, neutral excitons exhibit a continuous blueshift as temperature decreases, with no abrupt energy shifts across the phase transition. The lack of an additional energy shift may arise from additional absorbents, e.g., PDMS residuals and water, at the interface that reduce strain coupling in such PDMS-stacked heterostructures.

Beyond pronounced energy transfer, we also identify charge transfer processes at the TMD/FTBQS interface. A local graphite back-gated structure was fabricated for TMD/FTBQS structures with and without the thin hBN spacer, as shown in the schematic in Fig. 5a. The graphite back-gate can modulate the doping levels of the heterostructures across the thick hBN gate dielectric. By monitoring the gate-dependent exciton responses in the TMD layer, we can estimate the type of charge transfer due to the interlayer interactions. Figure 5b shows a microscope image of the device used for this measurement. The fabricated device was designed to enable independent measurement of the $WSe_2$, $WSe_2$/FTBQS, and $WSe_2$/hBN/FTBQS heterostructure regions. Figure 5c and d show the gate-dependent PL spectra comparison of the monolayer $WSe_2$ and $WSe_2$/FTBQS heterostructure at 4 K. In Fig. 5c, the highest energy peak near 0V is the neutral exciton peak. With further *n*-doping (positive gate voltages) or *p*-doping (negative gate voltages), the neutral exciton decreases while the *n*- and *p*-type trions appear at lower energies with phonon replicas and dark excitons. Thus, using the onset of trions and other charged states in the TMD , we can track the changes in doping levels at the heterostructures.

As shown in Fig. 5d, for the same measurements performed on the $WSe_2$/FTQBS region, the onset of $n$-type trion near 1.7 eV, as marked by the horizontal dashed line, is shifted upwards, in comparison to Fig. 5c. This indicates that the molecular layer provides additional hole doping into the $WSe_2$ layer. Based on the amount of $WSe_2$ neutral level shift and the gate parallel plate capacitance model, we can estimate the amount of $p$-doping from the $WSe_2$ is around $3 \times 10^{11}/cm^{-2}$. We note that such slight hole doping from the FTBQS layer was consistently observed from room temperature to 4 K and did not show obvious variation with the HS and LS states of Fe(II). When applying negative gate voltages and injecting $p$-type carriers into the heterostructure, the heterostructure PL showed a weaker gate voltage dependence, with persisting neutral exicotn and no obvious $p$-type trions at higher doping compared to the monolayer $WSe_2$ region. We infer that Fermi-level pinning likely arises from surface impurities, dangling bonds, and vacancies in the molecular layers and interface. Such Fermi-level pinning is less significant (see Supplemental Fig. S6) at higher temperatures, consistent with the expectation of easier thermal activation of trapped states and defects in the molecules. On the other hand, the doping level shift remains at higher temperatures, indicating its origin in electronic band alignments. We also examine the adjacent regions where the thin hBN spacer is present. As shown in Fig. 5e and 5f for the gate-dependent PL of $WSe_2$/hBN and $WSe_2$/hBN/FTBQS, the thin hBN layer effectively blocks any charge transfer process, and their gate dependence is identical, while the energy transfer process quenches the exciton amplitude as discussed above.

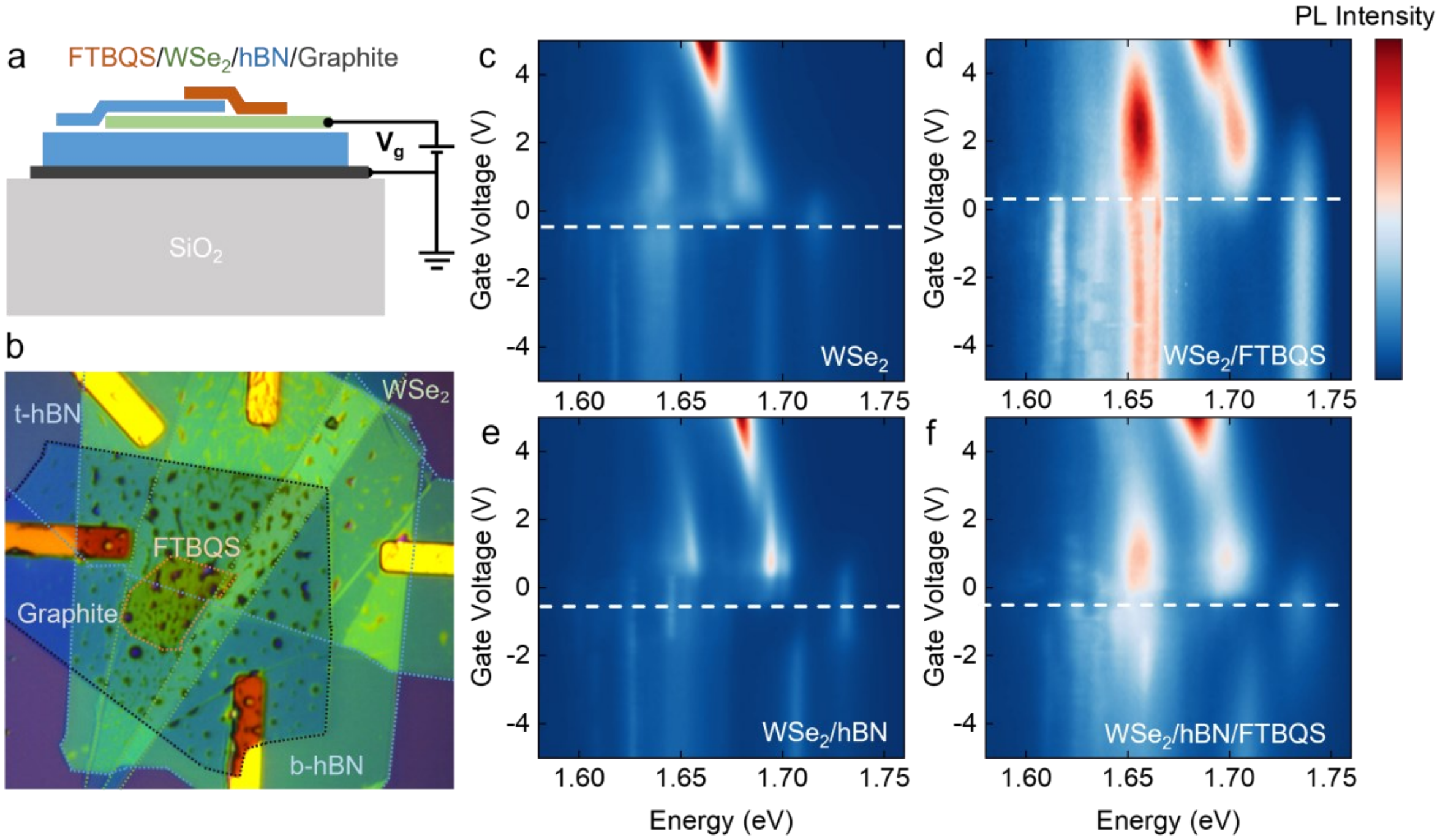

Figure 5. (a) Schematic of the heterostructure used for Gate-dependent PL measurements. The colors indicate the different layers. (b) microscope image of the same heterostructure, including regions where $WSe_2$, $WSe_2$/FTBQS, and $WSe_2$/hBN/FTBQS heterostructures can be measured independently. Gate-dependent PL spectra for the (c) $WSe_2$, (d) $WSe_2$/FTBQS, (e) $WSe_2$/hBN, and (f) $WSe_2$/hBN/FTBQS heterostructures at 4 K. The white dashed line indicates the charge neutral point.

The charge transfer process in molecular heterostructures with monolayer $MoS_2$, which has a very different work function[50] compared to $WSe_2$, is also examined to evaluate the impact of the different band alignment.

Due to the higher defect densities in $MoS_2$ crystals, its cryogenic PL spectra are dominated by defect-related emission. Here, we measured gate-dependent reflection contrast spectroscopy, and thus the exciton absorption features, at room temperature to identify the impact of charge transfer. We again compare the regions of $MoS_2$, $MoS_2$/FTBQS, $MoS_2$/hBN, and $MoS_2$/hBN/FTBQS. In Fig. 6a, the reflection contrast of $MoS_2$ is measured as a function of the back-gate voltage. The Si/$SiO_2$ substrate introduces additional interference signals in the reflection contrast, but we can clearly identify the emergence of neutral exciton signals in $MoS_2$ near 1.88 eV with decreasing gate voltage, as exfoliated $MoS_2$ is known to be intrinsically $n$-doped. In comparison, the $MoS_2$/FTBQS region appears more heavily $n$-doped, as the neutral exciton reflection contrast peak emerges and becomes pronounced at more negative back-gate voltages. The additional $n$-type doping from the molecular interface is estimated to be ~ $2 \times 10^{12}/cm^2$. Measurements of the comparison set with the thin hBN spacer again show that blocking the charge transfer with the spacer results in the same doping level and similar gate dependence (Fig. 6c&d), confirming the origin of this doping level shift.

The changes in charge transfer type in $WSe_2$/FTBQS and $MoS_2$/FTBQS can be understood by considering their work function and relative band alignments. The opposite charge transfer indicates different electronic band alignment between the TMD and FTBQS. As shown in the band schematic in Fig. 6e, we infer that FTBQS forms a staggered type-II band alignment with both $WSe_2$ and $MoS_2$, which is also consistent with the calculated LUMO-HOMO band gap[49] and the known work functions[50] of $WSe_2$ and $MoS_2$. Such a band alignment will favor a $p$-type charge transfer into $WSe_2$ and an $n$-type charge transfer into $MoS_2$, aggreging with our experimental observation. In addition, we note that the changes in doping level across the heterostructure regions are relatively homogeneous. This contrast in different charge transfer types also supports that the observed doping changes are due to the alignment of electronic bands rather than extrinsic effects, such as local defects, dopants, and dangling bonds at the molecular surfaces, which tend to provide the same type of doping to the adjacent layers. We note that the bands formed by the highest occupied and lowest unoccupied molecular orbitals of FTBQS mainly consist of the d orbital states of Fe[49]. The presence of interlayer charge transfer, mediated by tunneling/hopping processes, indicates significant electronic wave function overlap between the TMD and FTBQS layers despite the relatively large molecular size (~1.7nm), interlayer distance, and the tert-butyl groups. We did not observe significant changes in charge transfer rates across the SCO transition when varying the sample temperature, as expected, since the Fe d electrons lack spontaneous magnetization in the HS and LS states. The observation of charge transfer between such magnetic molecules and 2D semiconductors establishes future electrical pathways towards molecular properties modulation.

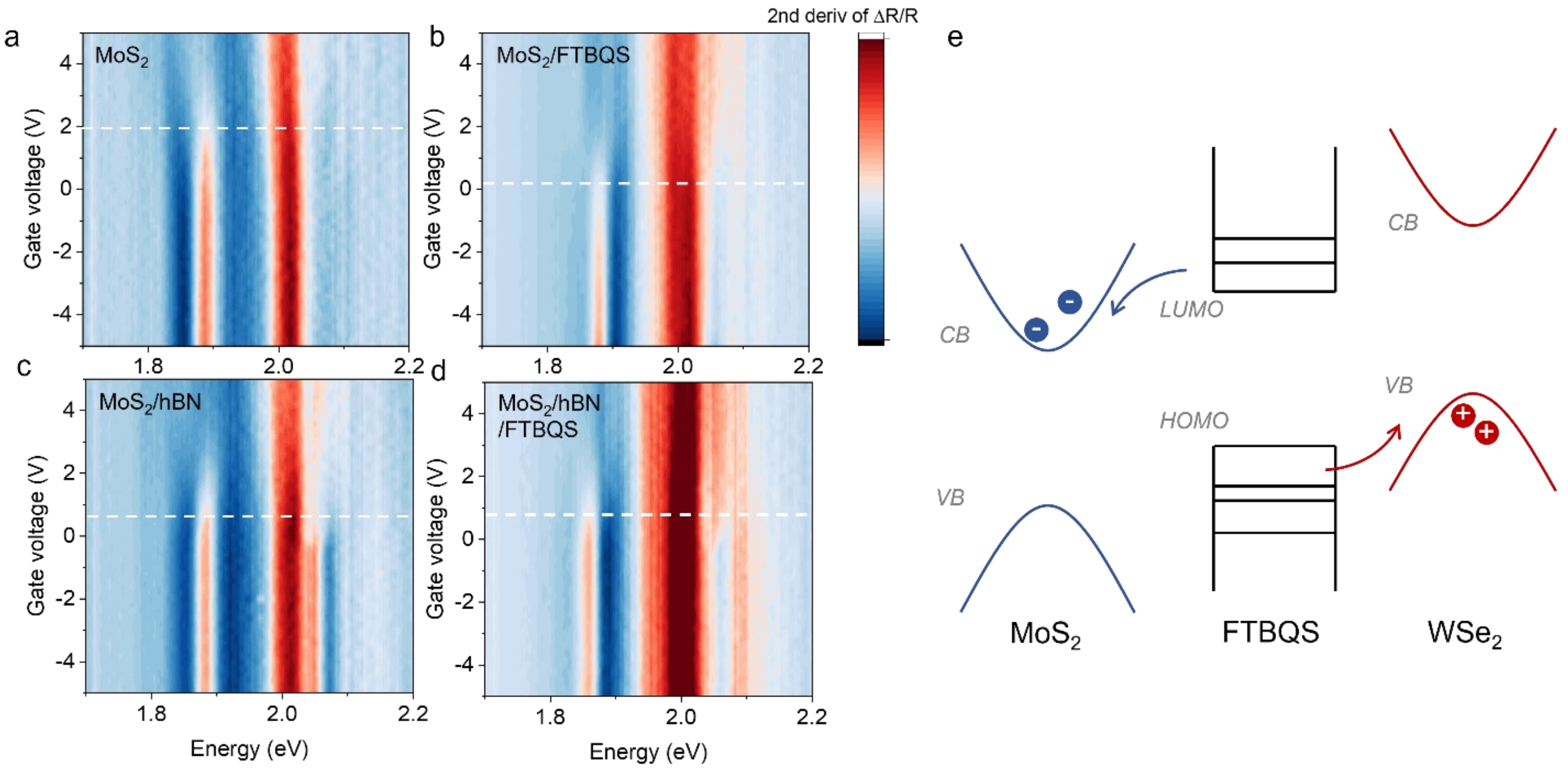

Figure 6. Comparison of the 2nd order derivative (to energy) of the reflection contrast $\Delta R/R$ spectroscopy of $MoS_2$ (a), $MoS_2$/FTBQS (b), $MoS_2$/hBN (c), $MoS_2$/hBN/FTBQS (d). The gate-dependent feature around 1.88 eV corresponds to the A exciton absorption. The white dashed lines are used to indicate their different Fermi level. (e) Schematic for the inferred band alignments between the valence (VB) and conduction bands (CB) of the monolayer $MoS_2$ (blue), $WSe_2$ (red) and the LUMO-HOMO bands of FTBQS (black). Data taken at room temperature.

In summary, we demonstrate strong interlayer interactions, both long-range energy transfer and short-range charge transfer, in non-planar molecule/2D material heterostructures. In particular, the long-range energy transfer shows the scaling law consistent with 2D systems and persists beyond 10 nm, enabling efficient coupling with complex, discrete molecules. These properties open up opportunities to build hybrid heterostructures that manipulate and detect a wide range of molecular electronic, optical, and spin properties with nanoscale solid-state devices.

AUTHOR INFORMATION

**Corresponding Author**

*Xiao-Xiao Zhang, xxzhang@ufl.edu

**Author Contributions**

X.-X.Z. conceived the experiments and supervised the device fabrication and optical measurements. M.G.-T. and M.S. synthesized the bulk molecular crystals. J.A. and J.K.-P. fabricated the devices and performed optical measurements. J.A. and K.H. analyzed the data and contributed to data interpretation. T.T. and K.W. synthesized the hBN crystals. K.H. and X.-X.Z. wrote the manuscript with input from all authors.

**Supporting Information**

Supplementary Figure Fig. S1-S6.

**REFERENCES**

1 Jariwala, D., Marks, T. J. & Hersam, M. C. Mixed-dimensional van der Waals heterostructures. *Nature Materials* **16**, 170-181, doi:10.1038/nmat4703 (2017).

2 Liu, Y., Huang, Y. & Duan, X. Van der Waals integration before and beyond two-dimensional materials. *Nature* **567**, 323-333, doi:10.1038/s41586-019-1013-x (2019).

3 Sato, O., Tao, J. & Zhang, Y.-Z. Control of Magnetic Properties through External Stimuli. *Angewandte Chemie International Edition* **46**, 2152-2187, doi:https://doi.org/10.1002/anie.200602205 (2007).

4 Zakrzewski, J. J., Liberka, M., Wang, J., Chorazy, S. & Ohkoshi, S.-i. Optical Phenomena in Molecule-Based Magnetic Materials. *Chemical Reviews* **124**, 5930-6050, doi:10.1021/acs.chemrev.3c00840 (2024).

5 Horiuchi, S. & Tokura, Y. Organic ferroelectrics. *Nature Materials* **7**, 357-366, doi:10.1038/nmat2137 (2008).

6 Dragulescu-Andrasi, A. *et al.* Radical Dimerization in a Plastic Organic Crystal Leads to Structural and Magnetic Bistability with Wide Thermal Hysteresis. *Journal of the American Chemical Society* **141**, 17989-17994, doi:10.1021/jacs.9b09533 (2019).

7 Lekin, K. *et al.* Heat, Pressure and Light-Induced Interconversion of Bisdithiazolyl Radicals and Dimers. *Journal of the American Chemical Society* **136**, 8050-8062, doi:10.1021/ja502753t (2014).

8 Gaita-Ariño, A., Luis, F., Hill, S. & Coronado, E. Molecular spins for quantum computation. *Nature Chemistry* **11**, 301-309, doi:10.1038/s41557-019-0232-y (2019).

9 Joachim, C., Gimzewski, J. K. & Aviram, A. Electronics using hybrid-molecular and mono-molecular devices. *Nature* **408**, 541-548, doi:10.1038/35046000 (2000).

10 Chen, X. *et al.* Efficient energy transfer in a hybrid organic-inorganic van der Waals heterostructure. *Science Advances* **11**, eadw3969, doi:10.1126/sciadv.adw3969 (2025).

11 Chowdhury, T. *et al.* Spectra-Orthogonal Optical Anisotropy in Wafer-Scale Molecular Crystal Monolayers. *Nano Letters* **25**, 5852-5859, doi:10.1021/acs.nanolett.5c00731 (2025).

12 Jariwala, D. *et al.* Hybrid, Gate-Tunable, van der Waals p–n Heterojunctions from Pentacene and MoS2. *Nano Letters* **16**, 497-503, doi:10.1021/acs.nanolett.5b04141 (2016).

13 Bennecke, W. *et al.* Hybrid Frenkel–Wannier excitons facilitate ultrafast energy transfer at a 2D–organic interface. *Nature Physics* **21**, 1973-1980, doi:10.1038/s41567-025-03075-5 (2025).

14 Chowdhury, T. *et al.* Emergent Above-Gap Photoluminescence in Molecularly Engineered Hybrid Bilayer Crystals. *ACS Nano* **19**, 40892-40901, doi:10.1021/acsnano.5c12187 (2025).

15 Zhu, T. *et al.* Highly mobile charge-transfer excitons in two-dimensional WS2/tetracene heterostructures. *Science Advances* **4**, eaao3104, doi:10.1126/sciadv.aao3104.

16 Ernst, K.-H. Molecular chirality at surfaces. *physica status solidi (b)* **249**, 2057-2088, doi:https://doi.org/10.1002/pssb.201248188 (2012).

17 Yang, S.-H., Naaman, R., Paltiel, Y. & Parkin, S. S. P. Chiral spintronics. *Nature Reviews Physics* **3**, 328-343, doi:10.1038/s42254-021-00302-9 (2021).

18 Naaman, R., Paltiel, Y. & Waldeck, D. H. Chiral molecules and the electron spin. *Nature Reviews Chemistry* **3**, 250-260, doi:10.1038/s41570-019-0087-1 (2019).

19 Rinehart, J. D. & Long, J. R. Exploiting single-ion anisotropy in the design of f-element single-molecule magnets. *Chemical Science* **2**, 2078-2085, doi:10.1039/c1sc00513h (2011).

20 Goodwin, C. A. P., Ortu, F., Reta, D., Chilton, N. F. & Mills, D. P. Molecular magnetic hysteresis at 60 kelvin in dysprosocenium. *Nature* **548**, 439-442, doi:10.1038/nature23447 (2017).

21 Guo, F.-S. *et al.* Magnetic hysteresis up to 80 kelvin in a dysprosium metallocene single-molecule magnet. *Science* **362**, 1400-1403, doi:10.1126/science.aav0652 (2018).

22 Atzori, M. & Sessoli, R. The Second Quantum Revolution: Role and Challenges of Molecular Chemistry. *Journal of the American Chemical Society* **141**, 11339-11352, doi:10.1021/jacs.9b00984 (2019).

23 Gong, C. & Zhang, X. Two-dimensional magnetic crystals and emergent heterostructure devices. *Science* **363**, eaav4450, doi:10.1126/science.aav4450 (2019).

24 Wasielewski, M. R. *et al.* Exploiting chemistry and molecular systems for quantum information science. *Nature Reviews Chemistry* **4**, 490-504, doi:10.1038/s41570-020-0200-5 (2020).

25 Mattia, E. & Otto, S. Supramolecular systems chemistry. *Nature Nanotechnology* **10**, 111-119, doi:10.1038/nnano.2014.337 (2015).

26 Amabilino, D. B., Smith, D. K. & Steed, J. W. Supramolecular materials. *Chemical Society Reviews* **46**, 2404-2420, doi:10.1039/c7cs00163k (2017).

27 Diao, Y., Shaw, L., Bao, Z. & Mannsfeld, S. C. B. Morphology control strategies for solution-processed organic semiconductor thin films. *Energy & Environmental Science* **7**, 2145-2159, doi:10.1039/c4ee00688g (2014).
28 Diao, Y. *et al.* Solution coating of large-area organic semiconductor thin films with aligned single-crystalline domains. *Nature Materials* **12**, 665-671, doi:10.1038/nmat3650 (2013).
29 Thompson, J. J. P., Gerhard, M., Witte, G. & Malic, E. Optical signatures of Förster-induced energy transfer in organic/TMD heterostructures. *npj 2D Materials and Applications* **7**, 69, doi:10.1038/s41699-023-00430-z (2023).
30 Candini, A., Klyatskaya, S., Ruben, M., Wernsdorfer, W. & Affronte, M. Graphene Spintronic Devices with Molecular Nanomagnets. *Nano Letters* **11**, 2634-2639, doi:10.1021/nl2006142 (2011).
31 Gavara-Edo, M. *et al.* Electrical Sensing of the Thermal and Light-Induced Spin Transition in Robust Contactless Spin-Crossover/Graphene Hybrid Devices. *Advanced Materials* **34**, 2202551, doi:https://doi.org/10.1002/adma.202202551 (2022).
32 Boix-Constant, C. *et al.* Strain Switching in van der Waals Heterostructures Triggered by a Spin-Crossover Metal–Organic Framework. *Advanced Materials* **34**, 2110027, doi:https://doi.org/10.1002/adma.202110027 (2022).
33 Dugay, J. *et al.* Phase Transitions in Spin-Crossover Thin Films Probed by Graphene Transport Measurements. *Nano Letters* **17**, 186-193, doi:10.1021/acs.nanolett.6b03780 (2017).
34 Gakiya-Teruya, M. *et al.* Asymmetric Design of Spin-Crossover Complexes to Increase the Volatility for Surface Deposition. *Journal of the American Chemical Society* **143**, 14563-14572, doi:10.1021/jacs.1c04598 (2021).
35 Koptur-Palenchar, J. *et al.* Thickness-dependent spin bistable transitions in single-crystalline molecular 2D material. *npj 2D Materials and Applications* **6**, 59, doi:10.1038/s41699-022-00335-3 (2022).
36 Selig, M., Malic, E., Ahn, K. J., Koch, N. & Knorr, A. Theory of optically induced F\"orster coupling in van der Waals coupled heterostructures. *Physical Review B* **99**, 035420, doi:10.1103/PhysRevB.99.035420 (2019).
37 Tebbe, D. *et al.* Distance Dependence of the Energy Transfer Mechanism in ${\mathrm{WS}}_{2}$-Graphene Heterostructures. *Physical Review Letters* **132**, 196902, doi:10.1103/PhysRevLett.132.196902 (2024).
38 Gaudreau, L. *et al.* Universal Distance-Scaling of Nonradiative Energy Transfer to Graphene. *Nano Letters* **13**, 2030-2035, doi:10.1021/nl400176b (2013).
39 Splendiani, A. *et al.* Emerging Photoluminescence in Monolayer MoS2. *Nano Letters* **10**, 1271-1275, doi:10.1021/nl903868w (2010).
40 Mak, K. F., Lee, C., Hone, J., Shan, J. & Heinz, T. F. Atomically Thin ${\mathrm{MoS}}_{2}$: A New Direct-Gap Semiconductor. *Physical Review Letters* **105**, 136805, doi:10.1103/PhysRevLett.105.136805 (2010).
41 Conley, H. J. *et al.* Bandgap Engineering of Strained Monolayer and Bilayer MoS2. *Nano Letters* **13**, 3626-3630, doi:10.1021/nl4014748 (2013).
42 Zipfel, J. *et al.* Spatial extent of the excited exciton states in ${\mathrm{WS}}_{2}$ monolayers from diamagnetic shifts. *Physical Review B* **98**, 075438, doi:10.1103/PhysRevB.98.075438 (2018).

43 Raja, A. *et al.* Energy Transfer from Quantum Dots to Graphene and MoS2: The Role of Absorption and Screening in Two-Dimensional Materials. *Nano Letters* **16**, 2328-2333, doi:10.1021/acs.nanolett.5b05012 (2016).
44 Lakowicz, J. R. *Principles of fluorescence spectroscopy*. (Springer, 2006).
45 Palummo, M., Bernardi, M. & Grossman, J. C. Exciton Radiative Lifetimes in Two-Dimensional Transition Metal Dichalcogenides. *Nano Letters* **15**, 2794-2800, doi:10.1021/nl503799t (2015).
46 Wang, G. *et al.* Colloquium: Excitons in atomically thin transition metal dichalcogenides. *Reviews of Modern Physics* **90**, 021001, doi:10.1103/RevModPhys.90.021001 (2018).
47 Rigosi, A. F., Hill, H. M., Li, Y., Chernikov, A. & Heinz, T. F. Probing Interlayer Interactions in Transition Metal Dichalcogenide Heterostructures by Optical Spectroscopy: MoS2/WS2 and MoSe2/WSe2. *Nano Letters* **15**, 5033-5038, doi:10.1021/acs.nanolett.5b01055 (2015).
48 Ye, Z. *et al.* Efficient generation of neutral and charged biexcitons in encapsulated WSe2 monolayers. *Nature Communications* **9**, 3718, doi:10.1038/s41467-018-05917-8 (2018).
49 Le, D., Jiang, T., Gakiya-Teruya, M., Shatruk, M. & Rahman, T. S. On stabilizing spin crossover molecule [Fe (tBu2qsal) 2] on suitable supports: insights from ab initio studies. *Journal of Physics: Condensed Matter* **33**, 385201 (2021).
50 Liu, Y., Stradins, P. & Wei, S.-H. Van der Waals metal-semiconductor junction: Weak Fermi level pinning enables effective tuning of Schottky barrier. *Science Advances* **2**, e1600069, doi:doi:10.1126/sciadv.1600069 (2016).